\documentclass{article}

\usepackage[preprint]{neurips_2021}

\usepackage{microtype}
\usepackage{graphicx}
\usepackage{subfigure}
\usepackage{booktabs} 
\usepackage{amssymb}
\usepackage{amsmath}
\usepackage{xcolor}
\usepackage{amsthm}
\usepackage{mathtools}

\usepackage[utf8]{inputenc} 
\usepackage[T1]{fontenc}    
\usepackage{hyperref}       
\usepackage{url}            
\usepackage{amsfonts}       
\usepackage{nicefrac}       

\def\lesssim{\mathrel{\hbox{\rlap{\hbox{\lower4pt\hbox{$\sim$}}}\hbox{$<$}}}}
\def\gtrsim{\mathrel{\hbox{\rlap{\hbox{\lower4pt\hbox{$\sim$}}}\hbox{$>$}}}}
\def\alt{\mathrel{\hbox{\rlap{\hbox{\lower4pt\hbox{$\sim$}}}\hbox{$<$}}}}
\def\agt{\mathrel{\hbox{\rlap{\hbox{\lower4pt\hbox{$\sim$}}}\hbox{$>$}}}}

\newcommand{\msun}{{\rm M}_{\odot}}
\newcommand{\beq}{\begin{equation}}
\newcommand{\eeq}{\end{equation}}
\newcommand{\bea}{\begin{eqnarray}}
\newcommand{\eea}{\end{eqnarray}}

\title{Interpreting a Machine Learning Model for Detecting  Gravitational Waves}

\author{%
  Mohammadtaher Safarzadeh\\
  Gravitational Astrophysics Laboratory, NASA Goddard Space Flight Center\\
  \texttt{mtsafarzadeh@gmail.com} \\
  \And 
  Asad Khan\\
  Department of Physics, University of Illinois at Urbana-Champaign\\
  Data Science and Learning Division, Argonne National Laboratory\\
  National Center for Supercomputing Applications, University of Illinois at Urbana-Champaign\\
  \AND
  E.~A. Huerta\\
  Data Science and Learning Division, Argonne National Laboratory\\
  Department of Computer Science, The University of Chicago\\
  \And
  Martin Wattenberg\\
  School of Engineering and Applied Sciences, Harvard University\\
}

\begin{document}

\maketitle

\begin{abstract}
We describe a case study of translational research, applying interpretability techniques developed for computer vision to machine learning models used to 
search for and find gravitational waves. The models we study are trained to detect black hole merger events in 
non-Gaussian and 
non-stationary advanced Laser Interferometer Gravitational-wave Observatory (LIGO) data. We produced 
visualizations of the 
response of machine learning models when 
they process advanced LIGO data that contains 
real gravitational wave signals, noise 
anomalies, and 
pure advanced LIGO noise. Our findings shed light on
the responses of individual neurons in these machine 
learning models.
Further analysis suggests that different parts of the 
network appear to specialize in local versus global 
features, and that this difference appears to be rooted 
in the branched architecture of the network as well as 
noise characteristics of the LIGO detectors. We believe 
efforts to whiten these ``black box'' models can suggest 
future avenues for research and help 
inform the design of interpretable machine 
learning models for gravitational wave astrophysics.  
\end{abstract}


\section{Introduction}
\label{sec:intro}

\noindent The advent of large scale 
scientific facilities~\citep{ska_ieee,ApollinariG.:2017ojx,lsstbook,HEPSoftwareFoundation:2017ggl}, 
next generation imaging technology, and 
supercomputing have enabled remarkable 
advances in contemporary science and engineering~\citep{gropp_foster,Huerta_big_data,conte_foster_gropp}. 
As these facilities continue to produce scientific datasets 
with ever increasing precision, volume and velocity, 
researchers have been empowered 
to make precision-scale measurements 
and predictions that are revolutionizing multiple
fields: cosmology, agriculture, personalized 
nutrition, genomics and particle physics~\citep{hep_kyle,Nat_Rev_2019_Huerta,Narita2020ArtificialIP,ai_agriculture,Uddin2019ArtificialIF,fair_hbb_dataset}.

Machine learning (ML) has begun to play an important 
role in this data-rich environment.
Recent successes have ranged from processing 
gravitational wave 
data and identify true gravitational wave signals in 
advanced LIGO data~\citep{geodf:2017a,George:2017vlv,george_huerta_plb,vizgws} to automating the classification of 
galaxies observed by the Dark Energy Survey and the Sloan Digital Sky Survey~\citep{viztsne2,viztsne,asad:2018K}, 
and star cluster images 
by Hubble~\citep{2020MNRAS.493.3178W,2021MNRAS.506.5294W}, 
to mention 
a few. Neural networks have been a key ML technique used in these projects.

Although neural networks have had many successes, 
they also have a key drawback when seen as scientific 
instruments: they often appear as black boxes, since how 
they make their decisions is unclear. Understanding 
how these neural networks work would have many 
general benefits~\citep{doshi2017towards}. Information 
on how neural networks function can help find 
bugs, inspire potential 
improvements~\citep{alsallakh2021debugging}, and help 
calibrate trust in their decisions \citep{carvalho2019machine}. 
In the context of astrophysics, there is also the potential 
that examining the features that a network thinks is important 
can give clues to new science \citep{2021Ntampaka}.

These questions have been investigated in many other 
contexts, ranging from image 
classification~\citep{simonyan2013deep} to natural 
language processing~\citep{tenney2019bert}. Our goal here 
is to describe a case study in interpretability, showing 
how scientists can apply techniques from this literature 
to a real-world scientific ML system, aimed at a very 
different domain. Our hope is to demonstrate 
the value of these techniques generally in the 
context of applications that involve complex, 
noisy, heterogeneous and incomplete scientific datasets.

In this article, we use gravitational wave 
astrophysics as a science 
driver to explore the inner workings of four
different neural network models that have been used 
for production scale, 
data-driven gravitational wave detection in 
advanced LIGO data~\citep{geodf:2017a,George:2017vlv,george_huerta_plb,vizgws,huerta_nature_ast}. We will refer to these models as the \textbf{GW ensemble.}

Our findings on the GW ensemble fall into two main 
categories. First, we find reassuring evidence that the 
ML model is based on physically meaningful features. 
A potential failure mode for an ML model is to pick up 
on confounding variables--for example, some medical 
image analysis systems have learned to identify the 
hospital and department where an image 
was taken~\citep{zech2018confounding}.
The fact that the GW model appears to focus on features 
that are intuitive to astrophysicists may increase trust 
in the system. Second, our explorations shed light on 
how the architecture of the GW model network may be 
influencing its computation. For example, the network 
is divided into two main branches, reflecting the 
physical architecture of the LIGO detectors. Our 
investigation suggests these branches have specialized 
in different ways.

While these results do not come close to a full explanation 
of the GW ensemble's decisions, they do provide a framework 
for understanding how those decisions are made, and can 
serve as a basis for future investigations of these 
models. More generally, our findings suggest that 
existing interpretability methods have reached the 
point where they may be broadly useful across domains.

\section{Detecting gravitational waves: model and data}
\label{sec:method}

We describe the ML models and data used in the GW ensemble, 
and the methods used to understand how 
these models distinguish between gravitational waves 
and noise anomalies. Because the domain is deeply 
technical, we begin with a basic description of 
the problem at hand, suitable for ML practitioners.

\subsection{Machine learning to detect black hole mergers}

Physicists have long been interested in 
``gravitational waves,'' ripples in the geometry of the 
universe caused by very compact objects, e.g.,  black holes or 
neutron stars, moving at a fraction of the speed of light. 
General relativity predicts the existence of these waves--and 
that they are extremely hard to detect. Confirming their 
existence was, for decades, a major unsolved problem in 
experimental physics. One particularly dramatic type of 
astronomical event is the merger of two black holes that 
will create ripples in space-time with large enough amplitude 
to be detected by detectors such as advanced LIGO. When two 
black holes merge, 
they orbit at decreasing distance and increasing speed, until 
they collide. The resulting gravitational waves should, 
theoretically, increase in magnitude and frequency until the 
point of collision--then cease.

Dramatic as these astronomical events are, they occur so far 
from the earth that the resulting waves are still weak compared 
to ambient noise. The LIGO detectors were designed to detect the 
presence of gravitational waves, and consist of two 
geographically separated sites, Hanford (Washington State) 
and Livingston (Louisiana). To confirm an observation of a 
binary black hole merger, physicists need to solve a 
basic classification problem: analyze time series from the Hanford 
and Livingston detectors, and classify whether the data 
includes a gravitational wave signal.

As a ML task, the problem can be abstracted as 
follows. The input consists of two vectors of real numbers, 
which correspond to time series from the two detectors for 
a small period of time. The output will be a vector of the 
same length, but consisting of numbers between 0 and 1, 
corresponding to a measurement of how likely it is that 
a black hole merger is happening at that instant. ML models 
for this task are typically trained on simulated waveforms 
(verified binary merger events are scarce) combined with real 
background noise (which is all too plentiful).

\subsection{Model and data}
\label{sec:data}

\noindent The ML ensemble we use in this study was 
introduced in~\cite{huerta_nature_ast}. As described 
below, these models share the same architecture and 
only differ in their weight initialization. The 
modeled waveforms we used to train them describe 
a 4-D signal manifold that encompasses the masses 
and individual spins of the binary components 
of  quasi-circular, spinning, non-precessing 
binary black hole mergers, namely, 
\((m_1, m_2, s^z_1, s^z_2)\). This is the same 
parameter space employed by 
were trained with modeled waveforms that describe the 
physics of quasi-circular, spinning, non-precessing 
binary black hole mergers, i.e., the same 4-D signal 
manifold of traditional, template-matching 
pipelines used for gravitational 
wave searches~\citep{2016CQGra..33u5004U,2018PhRvD..98b4050N,2021SoftX..1400680C}. 

These modeled waveforms were combined with advanced 
LIGO noise to expose these models to a diverse set of astrophysical 
scenarios encoding time and scale invariance, i.e., 
these ML models may identify gravitational waves irrespective 
of their signal to noise ratio or location in the 
input time-series advanced LIGO strain. Furthermore, 
these models were tested using hours-, weeks- and a 
month-long advanced LIGO dataset to quantify their 
ability to identify real events and discard noise 
anomalies. The studies presented in~\citet{huerta_nature_ast} 
showed that this ensemble of 4 ML models can 
concurrently process
data from Hanford and Livingston, 
correctly identifying all binary black hole mergers 
reported throughout August 2017 with no misclassifications. 

To enable other researchers to reproduce these 
studies, these AI models were released through 
the Data and Learning Hub for Science. To facilitate the 
use of these models we have demonstrated how to 
connect DLHub with the HAL GPU-cluster at the National 
Center for Supercomputing Applications, and do inference 
by using funcX as a universal computing service. 
Using the framework DLHub-funcX-HAL, we showed that 
it is possible to process a month's worth of advanced 
LIGO noise within 7 minutes using 64 NVIDIA V100 GPUs. 
It is worth contextualizing these results with other 
studies in the literature. For instance, the first 
class of neural networks for gravitational wave 
detection~\citep{geodf:2017a,George:2017vlv,george_huerta_plb} 
were tested using 4096 second long 
advanced LIGO data segments, reporting one misclassification 
for every 200 seconds of searched data. More 
sophisticated neural networks have been developed to 
test days- and weeks-long advanced LIGO data, reporting 
one misclassification for every 2.7 days of searched 
data~\citep{2021PhLB..81236029W}, and one misclassification 
for every month of searched 
data~\citep{2022arXiv220111133C}.
In brief, the ML ensemble we used in this study 
represents a class of neural networks that are 
adequate for accelerated, data-driven 
gravitational wave detection st scale.
The key components of our ML models and 
data encompass: 

\noindent \textbf{Modeled waveforms} Our 
ensemble of four ML models 
was trained with 1.1 
million waveforms produced with the 
\texttt{SEOBNRv3} approximant~\citep{seobnrv3}. 
These waveforms, sampled 
at 4096 Hz,  describe quasi-circular, spinning, non-precessing, 
binary black hole mergers with 
total mass \(M\in[5\msun,\,100\msun]\), 
mass-ratios \(q \leq 5\), and individual 
spins \(s^z_{\{1,2\}}\in[-0.8,\,0.8]\). These 1-second 
long waveforms describe the late-inspiral, merger 
and ringdown phases.

\noindent \textbf{Advanced LIGO noise} 
The modeled waveforms described above are whitened 
and linearly mixed with 4096-second 
long advanced LIGO Hanford and Livingston noise 
data segments obtained from the 
Gravitational Wave Open 
Science Center~\citep{Vallisneri:2014vxa}. The start 
GPS time of 
each of these segments is: $1186725888$, $1187151872$, and $1187569664$. None of these data 
segments contained gravitational wave signals.
The ground truth labels were encoded in a binary fashion such that time-steps after the 
merger are classified as noise [labelled 0], and 
all preceding time-steps in the 1 second long window 
are classified as waveform strain [labelled 1]. Hence 
the transition from 1s to 0s indicates the location of the merger. 

\begin{figure}[!ht]
\centerline{
\includegraphics[width=1.2\linewidth]{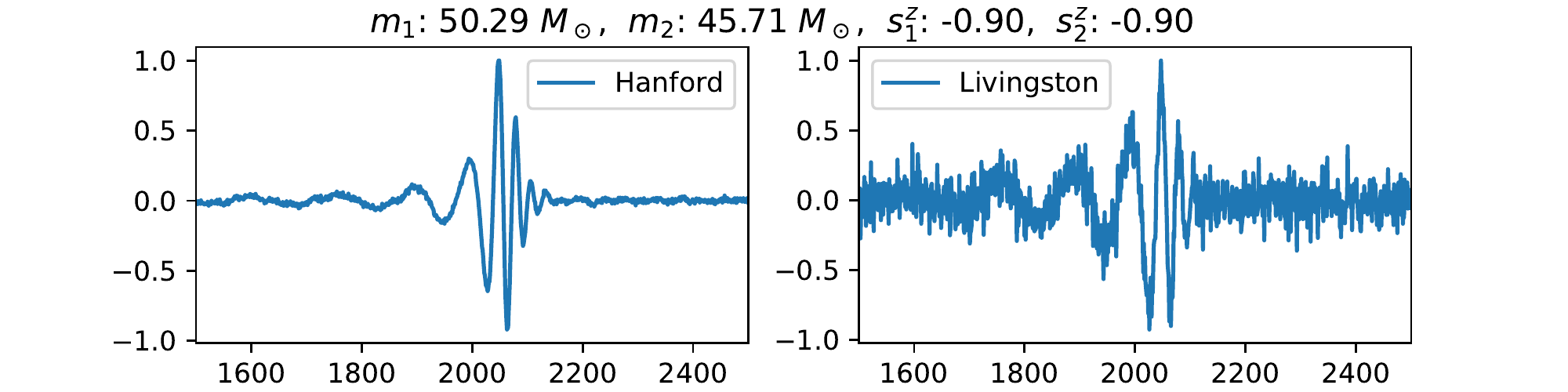}
} 
\caption{\textbf{Input data for training:} The 
panels show whitened waveforms with advanced LIGO noise 
using Hanford (left panel) and Livingston (right panel) 
strain data. The waveforms represent different configurations our neural networks 
are exposed to during the training stage.The example shown here is the GW of a 
a binary black hole with component masses 
\(m_1=50.29\msun, m_2=45.71\msun\) and individual spins 
\(s^z_1=s^z_2=-0.9\) that has been whitened with 
a PSD representative of O2 advanced LIGO noise and then 
linearly combined 
with whitened O2 noise.}
\label{fig:waveforms_in}
\end{figure}

To train the models, as described in~\citet{huerta_nature_ast},
co-authors of this article created a set of modeled 
waveforms, selected a batch of advanced LIGO noise to 
estimate a power spectral density (PSD), and then use this 
PSD to whiten both modeled waveforms and noise so that 
these may be linearly added at will to expose 
our AI models to modeled signals that 
are contaminated with multiple noise realizations 
and signal-to-noise ratios 
during training. Figure~\ref{fig:waveforms_in} presents 
a sample case in which a modeled waveform that describes 
a black hole binary with component masses 
\(m_1=50.29\msun, m_2=45.71\msun\) and individual spins 
\(s^z_1=s^z_2=-0.9\) has been whitened with 
a PSD representative of O2 advanced LIGO noise and then 
linearly combined 
with whitened O2 noise. At inference, our AI models process 
whitened advanced LIGO data at scale and produce 
a typical response function that differentiates 
real events from noise anomalies.

\subsection{Neural network architecture}
\label{sec:architecture}

\begin{figure}
    \centerline{
    \includegraphics[width=.5\linewidth]{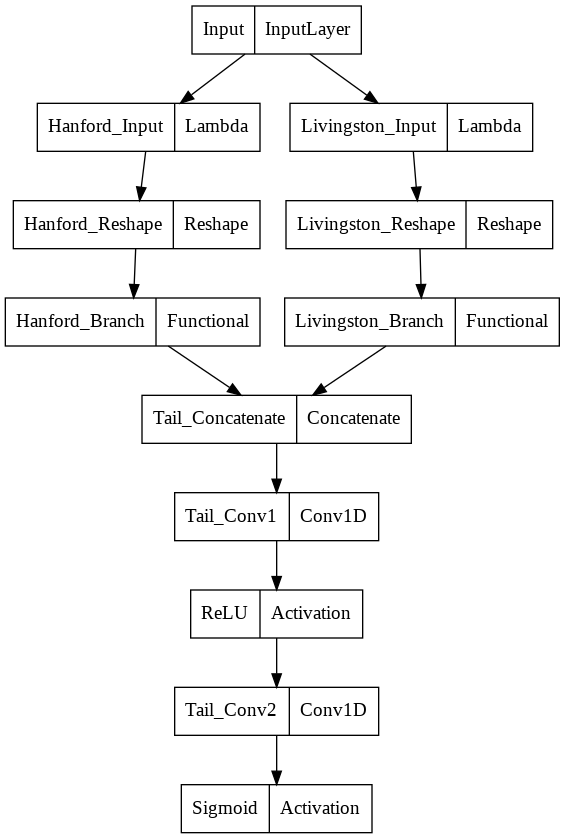}
    }
\caption{\textbf{Branched architecture:} This two branch
neural network structure is used to concurrently process 
Hanford and Livingston advanced LIGO data. The architecture of each branch closely follows that of \texttt{WaveNet} model ~\citep{2016wavenet}.}
\label{fig:model_architecture_branches}
\end{figure}

The architecture of each model in the GW ensemble 
consists of two branches and 
a shared tail, as shown in Figure~\ref{fig:model_architecture_branches}. The two 
branches are modified \texttt{WaveNet}s~\citep{2016wavenet} 
processing 
Livingston and Hanford strain data separately. The 
key components of a WaveNet architecture are dilated 
causal convolutions,  gated  activation  units, and the 
usage  of residual and skip  connections. Since ours is 
a  classification task instead of an autoregressive one, 
e.g., waveform generation, we turn off the causal padding 
in the convolutional layers. We use filter size of 2 in 
all convolutional layers and stack 3 residual blocks 
each consisting of 11 dilated convolutions so that 
the receptive field of each branch $> 4096$. The outputs 
from the two branches are then concatenated together 
and passed through two more convolutional layers 
of filter-size 1. Finally a \texttt{sigmoid} 
activation function is 
applied to ensure that the output values are in 
the range $[0,1]$. The complete network 
structure was introduced in~\citep{huerta_nature_ast}.

We trained multiple ML models with this architecture 
using 
the same training datasets (modeled waveforms \& advanced 
LIGO noise) but different weight initialization. We then 
selected 4 models out of our training suite (via a process to optimize ensemble accuracy) and 
wrote post-processing software that looks at the combined 
output of this ensemble.
We only label noise triggers 
as candidate events when the response of all four models 
is consistent with an event whose threshold for 
detectability is above a given threshold, as described in 
detail in~\citet{huerta_nature_ast}.

\section{Visualization and interpretation of the GW ensemble}
\label{sec:res}

The GW ensemble is a successful production system that 
has proven useful to scientists. As described above, it 
is accurate and efficient. Yet it is unclear how it 
achieves this success. The goal of our investigation is 
to show how techniques that have been developed for neural 
nets in other fields (primarily machine vision) can be used 
to shed light on different aspects of the ensemble.

We present three types of results, namely, 
sensitivity maps 
to peer into the response of our ML models to a 
variety of input data; scientific visualizations 
that showcase the response of our models to 
second-long input datasets; and 
activation maximization studies which 
suggest that individual neurons within the GW ensemble 
respond especially strongly to signals that resemble part or all of modeled merger events.

\subsection{Activation maximization}
\noindent \textbf{Background:} 

A natural question is whether individual neurons respond to characteristic types of signals. One way to investigate this question 
is to look for signals that maximally activate a 
particular neuron~\citep{erhan2009visualizing}. In the special 
case of a neuron that is computing a class score, this is 
roughly equivalent to finding an input signal that the 
network thinks is ``most likely'' to be of a certain class.

Conceptually, 
one may think of these maximally-activating signals as 
templates that the network is looking for. In the 
case of convolutional neural networks used in computer 
vision, one often sees a hierarchy of such templates. 
Lower-level neurons pick out simple signals, 
such as edges, and higher-level neurons seem to respond 
best to concepts such as an eye, or an entire 
dog~\citep{cammarata2020thread}. While this technique has 
largely been applied to 2D convolutional networks, although 
it has been applied to EEG data (via the frequency domain) 
in the case of \citet{ellis2021novel}.

\noindent \textbf{Application to time-series gravitational 
wave data:} 

To investigate the GW ensemble, we chose a 
middle position (element 2000) in the data window, and 
synthesized signals that maximize the neuron activations at this specific timestamp position for different layers. To understand aggregate activity, we 
maximized total activation in a given layer. We 
computed maximizing signals for individual neurons as well, 
which gave similar results.

We perform the maximization procedure 10 times for each 
layer. In all cases we have started 
with pure Gaussian noise with mean zero and standard deviation of 0.02. 
For all activation maximization studies in all of the layers 
we have adopted a learning rate of 0.5 with momentum 0.9 
using Adam optimizer for performing the  
calculations necessary to seek a maximizing signal. 
We use L = $\lvert 1- x_n \rvert^2$ as a loss, where $x_n$ is the value of the activation of neuron $n$, and stop when $L< 0.01$.

\noindent \textbf{Results:}

\begin{figure}
\centerline{
\includegraphics[width=\linewidth]{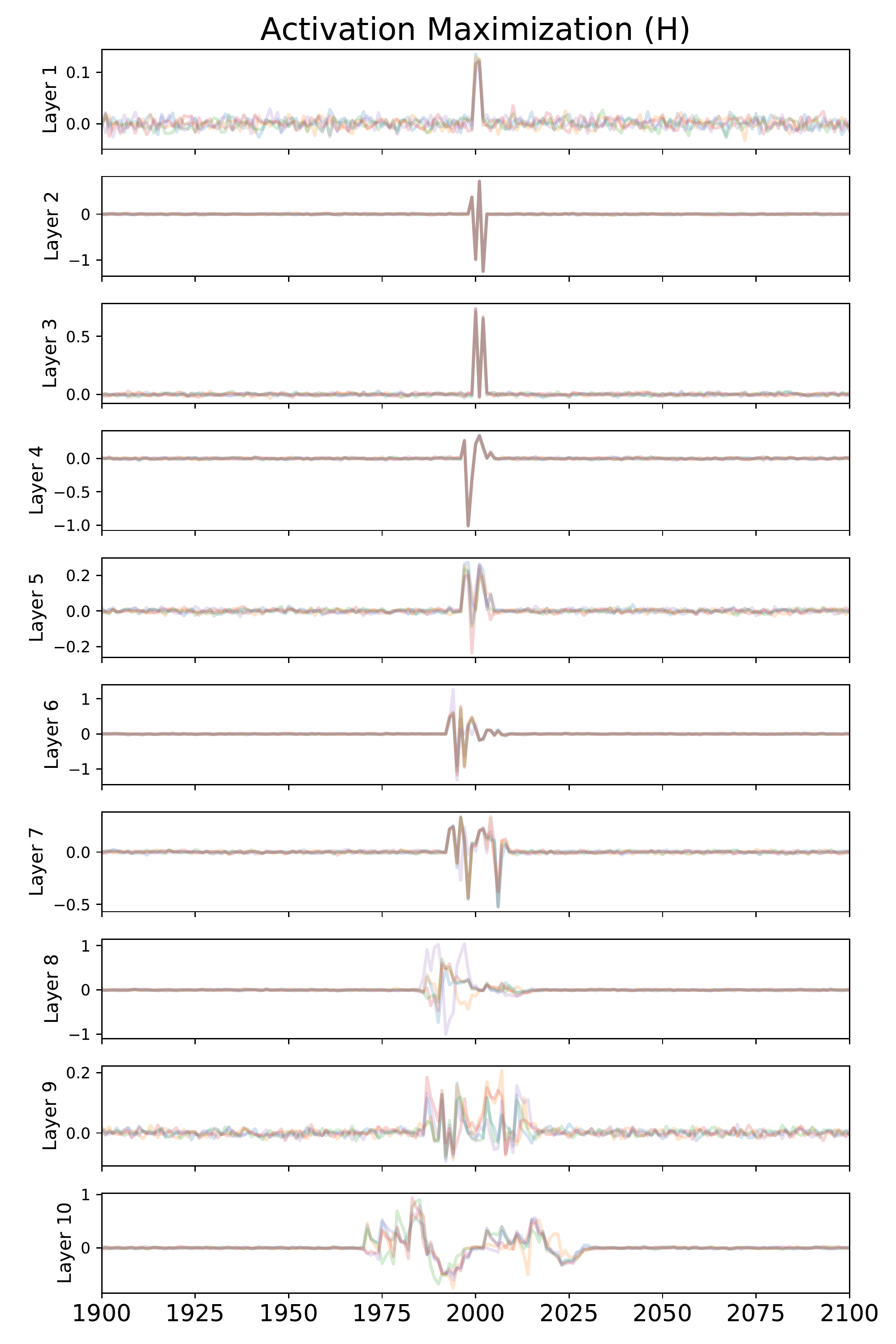}
} 
\caption{\textbf{Activation maximization in shallow layers:} 
Each panel presents how a single neuron's activation is 
maximized when we input random Gaussian noise with mean 
zero and standard deviation of 0.02. We repeat this procedure 
5 times, using learning rate of 0.1 and stochastic 
gradient descent optimizer with momentum 0.9.}
\label{fig:AM_layer_1_10}
\end{figure}

Results from the first 10 layers for Hanford branch are shown in 
Figure~\ref{fig:AM_layer_1_10}. We see that the 
maximally activating patterns are far from random noise, 
instead resembling different gravitational waveform-like 
signals. (Recall the training examples in Figure~\ref{fig:waveforms_in}.) Neurons in the first layer seem maximally 
activated by spikes, and as we move through deeper layers 
the signals become longer and more complex. 
The results are similar 
for the Livingston branch in these early layers. This 
pattern seems like a one-dimensional analogue of the pattern 
seen in networks aimed at computer 
vision\citep{erhan2009visualizing}.

\begin{figure}[!ht]
\centerline{
\includegraphics[width=1.2\linewidth]{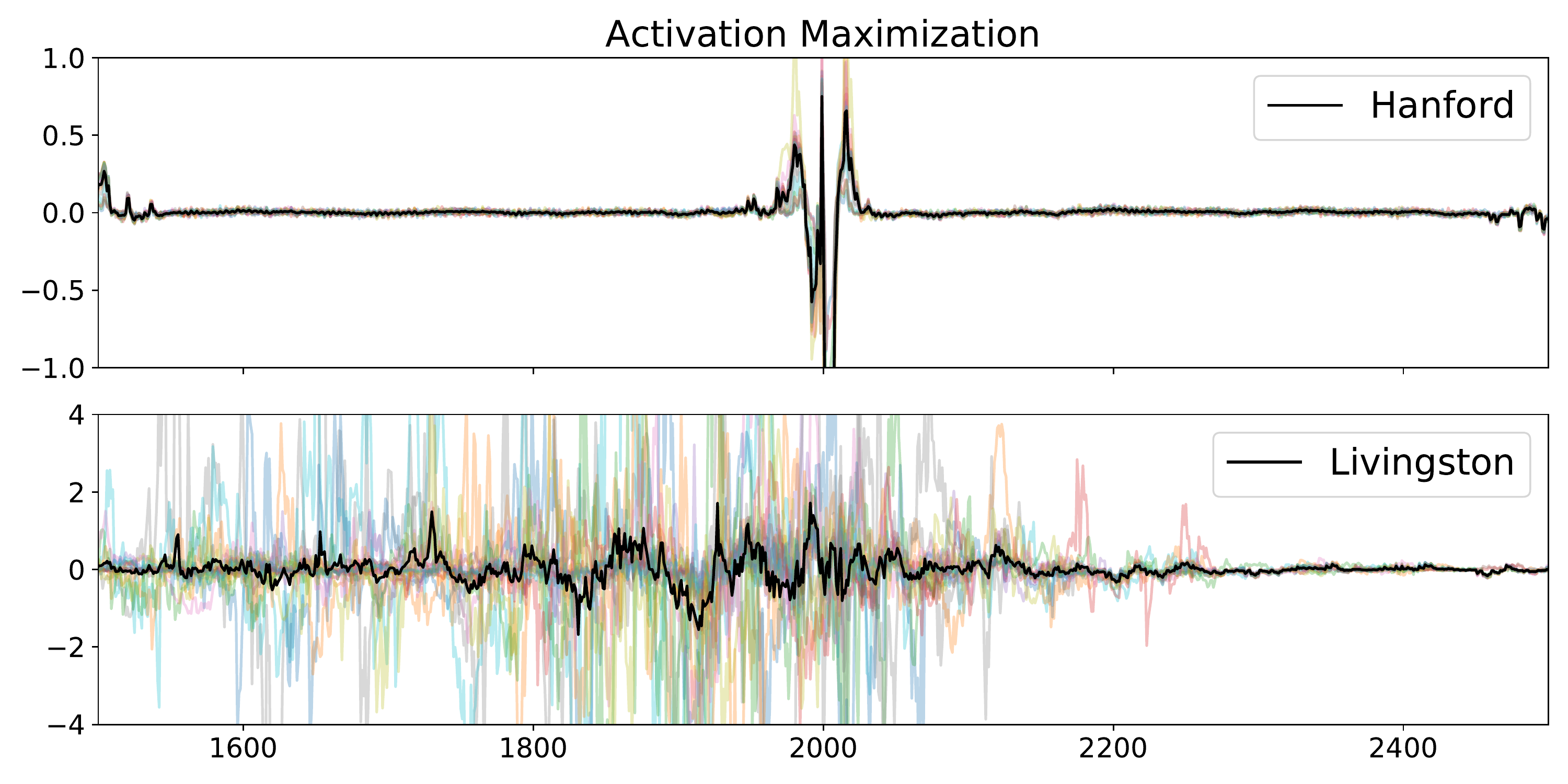}
} 
\caption{\textbf{Activation maximization in last layer:} Same 
as Figure~\ref{fig:AM_layer_1_10}, but now for the 
last activation layer of the network. The activation 
maximization is performed 20 times, the result of each is 
shown with a colored line. All the 20 tries have the 
initial condition of random Gaussian with mean zero and 
standard deviation of 0.02. The black line shows the average 
of the 20 tries. Top (bottom) panel shows the result for 
the Hanford (Livingston) branch of the network. (The difference between branches is a consistent theme in our analysis.) We have adopted a learning rate of 0.5 with momentum 0.9 using 
Adam optimizer for performing the gradient descent 
calculations. We refer the readers to the movie in this  \href{https://www.youtube.com/watch?v=SXFGMOtJwn0&feature=youtu.be}{\color{red}{\underline{YouTube link}}} showing the process of finding the activation maximizing signal from pure noise. }
\label{fig:AM_layer_last}
\end{figure}

Figure~\ref{fig:AM_layer_last} shows the activation 
maximizing signal for the last layer of the 
network. We notice that the result for the 
Hanford branch resembles a denoised, whitened waveform 
with the merger event very well defined. One possible interpretation of this analysis is that the 
the Hanford branch specializes in searching 
for and 
identifying local features that describe the merger event. 
On the other hand, the Livingston branch seems to react 
more sharply to noise contaminated waveform-like 
signals over extended periods of time, not just around 
the merger event. Potentially related aspects 
of the Hanford and Livingston branches
manifest themselves in similar fashion in the 
studies we present below.

\subsection{Visualizing activations}
 
\noindent \textbf{Method:}

A direct method of visualizing 
the response of a neural network is to show the activations 
of all neurons in a single image. In the case of the GW 
ensemble, this means displaying 12,914,688 activations in 
a single diagram.

While this may seem like an overwhelming number, if we 
allocate one pixel per neural unit, and use color to 
encode activation level, it becomes feasible to create 
a holistic view of the the response in every layer at once. 
There are two main questions that such a visualization 
can answer. One is, can we see any differences between 
how the network processes signals that contain a 
gravitational wave, versus one that is pure noise? 
The second is, are there systematic differences between 
the two branches of the network?

\noindent \textbf{Results:}

\begin{figure}[!ht]
\centerline{
\includegraphics[width=.6\linewidth]{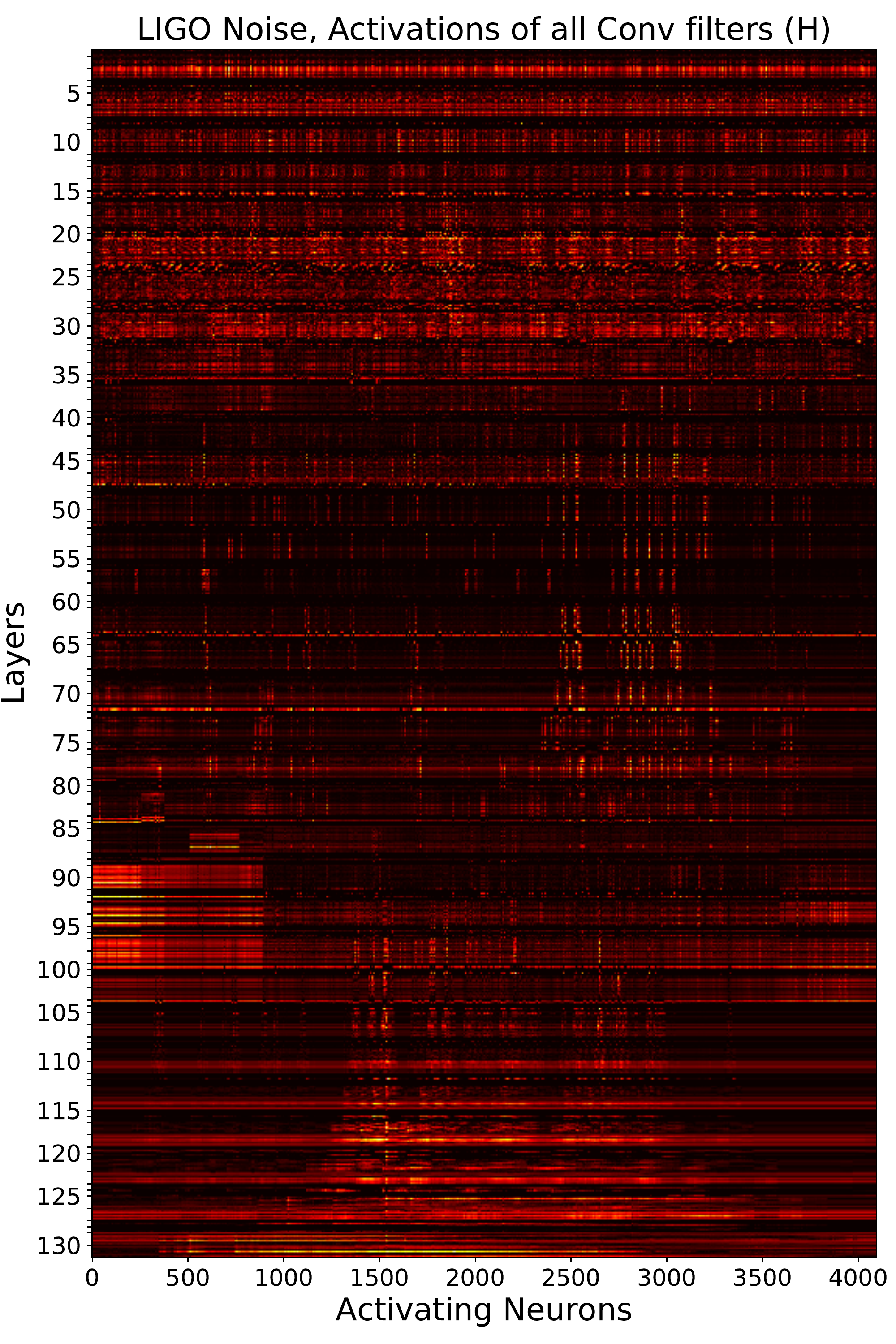}
\includegraphics[width=.6\linewidth]{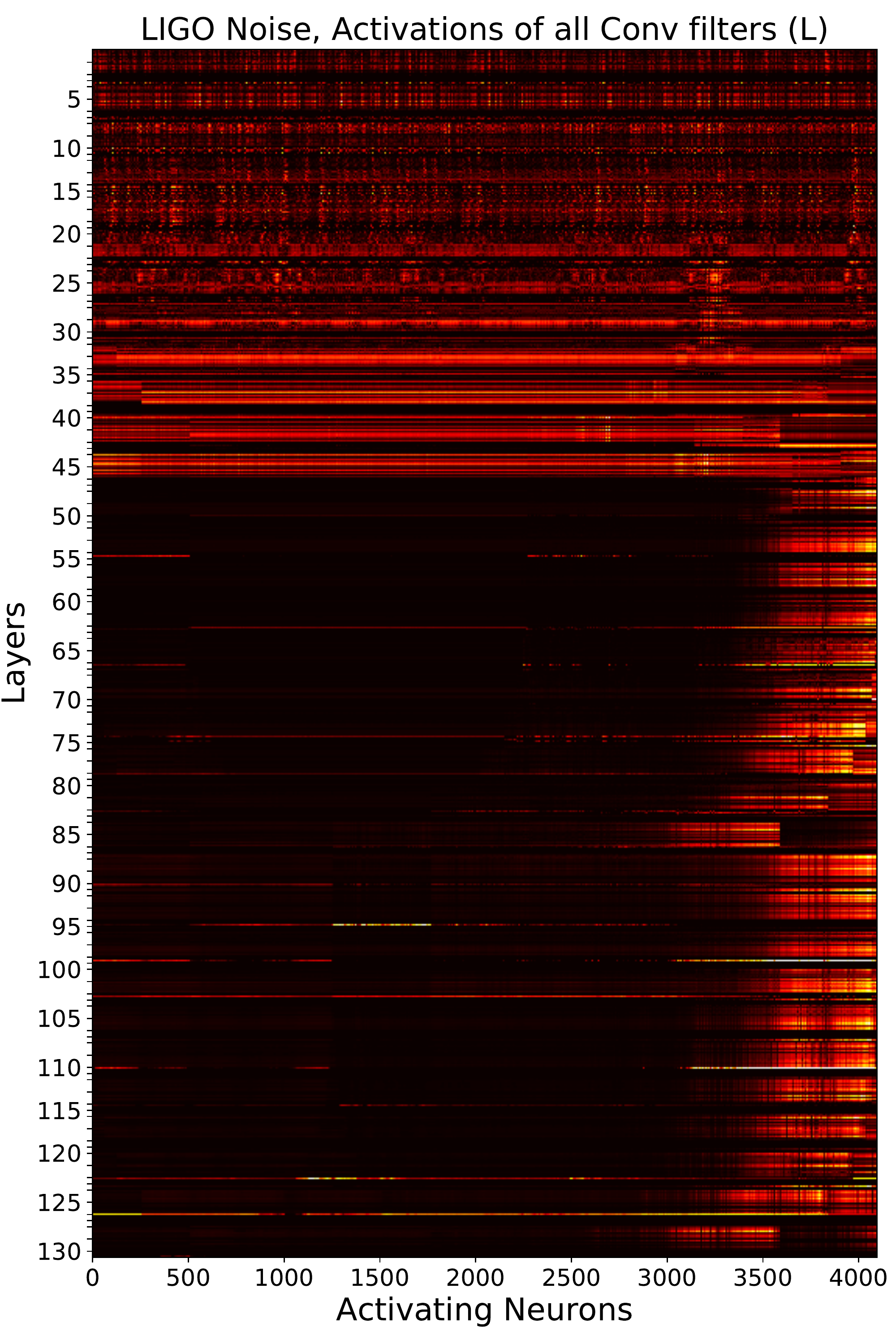}
} 
\caption{\textbf{Visualization of activations for pure noise:}
Each of our 
ML models has two branches that concurrently process 
Hanford and Livingstone strain data. These panels show the 
response of the convolutional layers in each of these 
networks from the shallow layers (top part) to the very last layers (deepest layers) for the Hanford  and Livingston 
branches. We used advanced LIGO data from the second observing run to produce this visualization.}
\label{fig:activations_noise}
\end{figure}

As a baseline, we begin by visualizing the response 
of the network to a non-merger event. The left panel 
in Figure~\ref{fig:activations_noise} shows 
the response of all the activation layers 
of one of the \texttt{WaveNet} models in the ensemble 
to advanced LIGO noise in the absence of signals.

\begin{figure}[!ht]
\centerline{
\includegraphics[width=.6\linewidth]{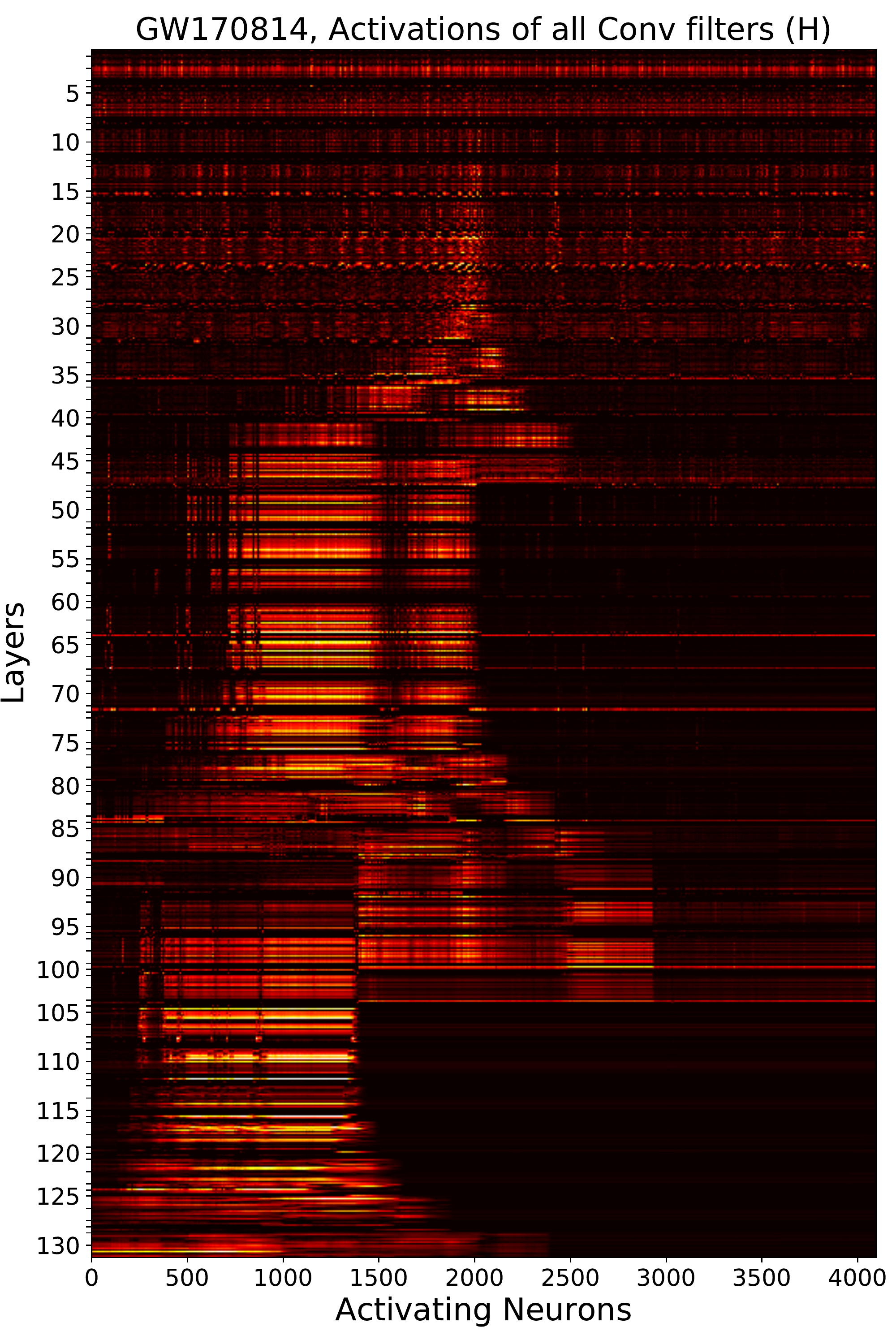}
\includegraphics[width=.6\linewidth]{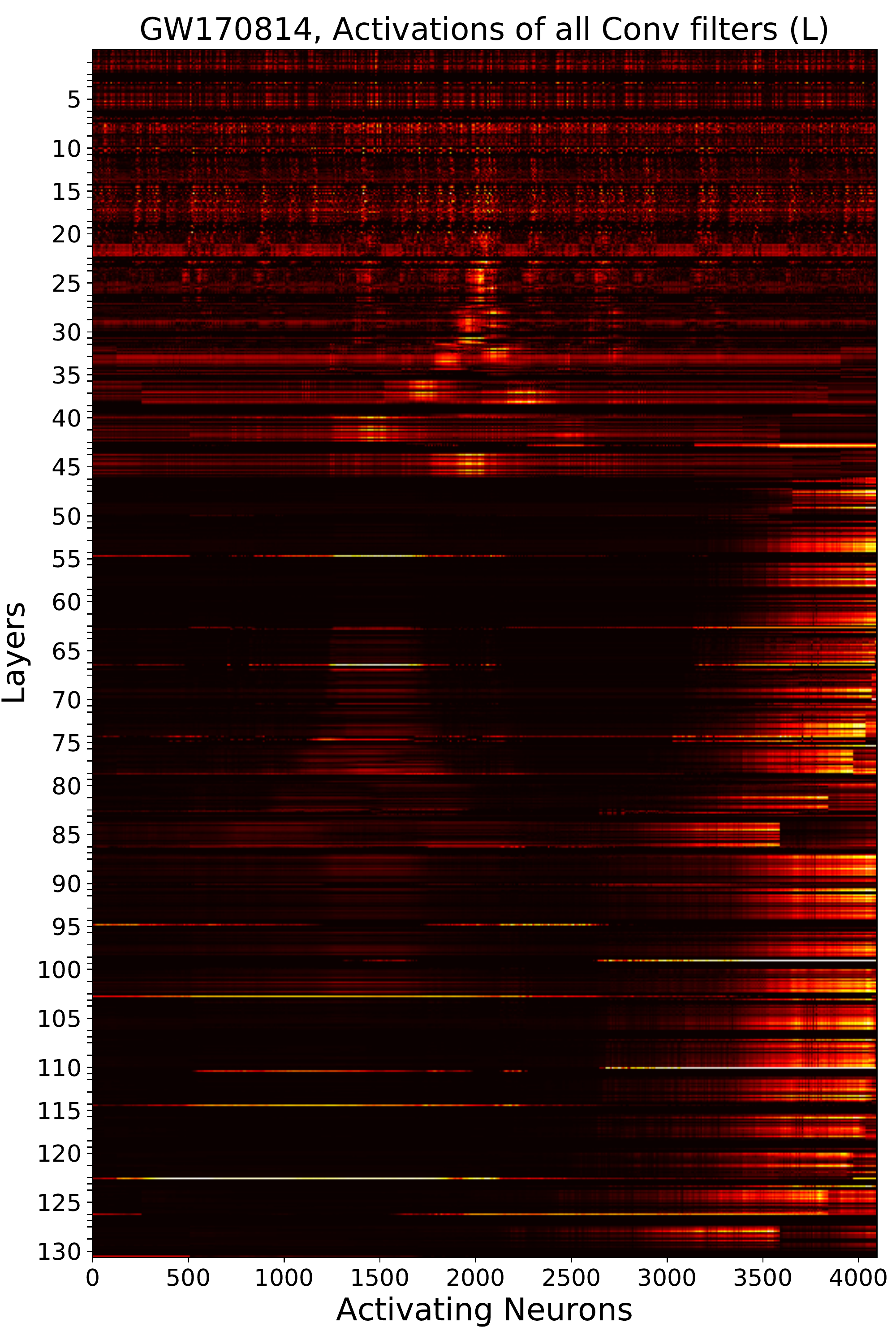}
} 
\caption{\textbf{Visualization of activations for GW170814:} Same as Figure~\ref{fig:activations_noise}, but now for a real gravitational wave source, GW170814. We notice a sharp, 
distinct response of this network for a signal whose merger is 
located around timestep 2000. We refer the readers to the movie in this  \href{https://www.youtube.com/watch?v=itVCj9gpmAs}{\color{red}{\underline{YouTube link}}} showing how neurons in different layers are  activated by the passage of a GW in the time series signal.}
\label{fig:activations_gw170814}
\end{figure}

Next, we conducted a similar analysis for the 
binary black hole merger GW170814, and present 
corresponding results in 
Figure~\ref{fig:activations_gw170814}. 
Pair-wise comparisons between 
Figures~\ref{fig:activations_noise} and~\ref{fig:activations_gw170814} show that our 
neural networks respond very differently, across all 
layers, when there is a signal in the data. The top 
layers show activations at the location of the merger event, 
gradually spreading out and branching as we move 
towards deeper layers. Although this result is not surprising, 
it provides some validation for the model and the 
visualization technique.

\begin{figure}[!ht]
\centerline{
\includegraphics[width=.6\linewidth]{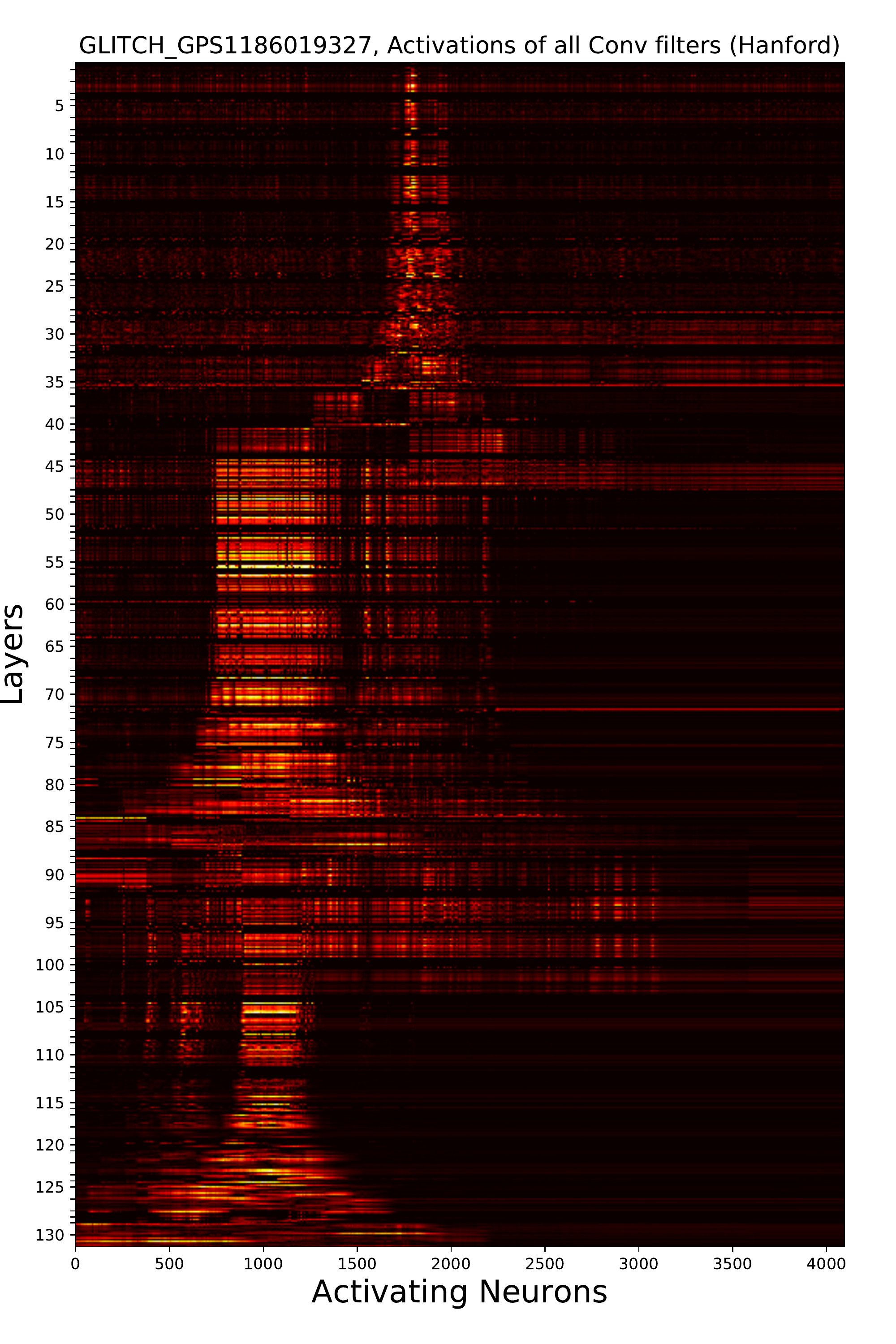}
\includegraphics[width=.6\linewidth]{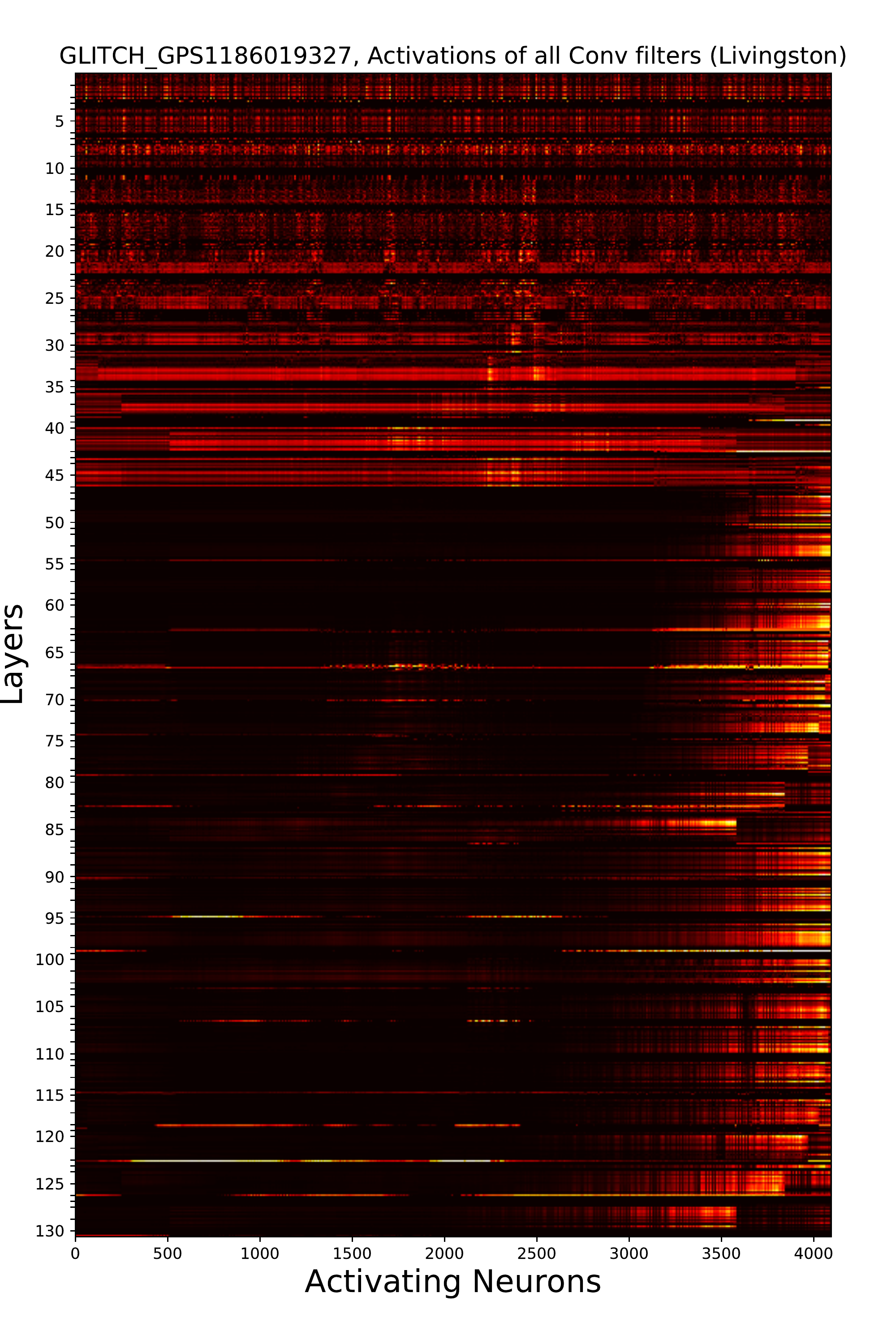}
} 
\caption{\textbf{Visualization of activations for a glitch} Unlike for the case of a real signal, the reaction of the network to a glitch is different in its top (early) layers: we observe a sense of fuzziness indicative of the uncertainly in the network as to where the merger event is taking place.
}
\label{fig:activations_glitch}
\end{figure}

As a final exploration, Figure~\ref{fig:activations_glitch} 
visualizes the response of the network to a ``glitch,'' i.e., 
a noise anomaly resembling a gravitational wave. The 
layers involved in the detection behave differently from 
when a real signal is present as shown 
in Figure~\ref{fig:activations_gw170814}. One obvious 
difference is the fuzziness of the activations in the 
early layers in detecting the merger event. One interpretation of this behavior is that it relates to the uncertainty of network as to 
where exactly the merger event is taking place.

We can make additional observations based on these 
visualizations: (i) a set of layers (from 20 through 45) in 
the Livingston branch remain very active, though with 
rather different behaviour, to all inputs 
we consider, i.e., pure noise, real signal and glitch, as 
shown in Figures~\ref{fig:activations_noise}-
\ref{fig:activations_glitch}, respectively. What we learn 
from these results is that the Livingston branch seems to look for 
global features in the input, in  
similarity to the findings we reported above on 
activation maximization; (ii) the response of the 
Hanford branch to noise anomalies or gravitational wave signals 
is sharper and more concentrated than that of the 
Livingston branch, as shown by layers 5 through to 35 in Figures~\ref{fig:activations_gw170814} and \ref{fig:activations_glitch}. Again, this is consistent with 
our findings above in which the Hanford 
branch seems to be maximally activated by more
localized features.

\subsection{Sensitivity Maps}

\noindent \textbf{Background:}

A basic technique for understanding the 
features relevant to a classification decision is 
to examine the gradient of the class scores, or equivalently the Jacobian of the final layer. Conceptually, 
the gradient tells us, for each feature, how much an 
infinitesimal change in that feature will affect the 
class score. A common hypothesis is that if the network 
is particularly sensitive to a given attribute, then 
that attribute is playing an important role in the 
classification. This technique, along with close variants, 
has been widely used in the machine vision community~\citep{simonyan2013deep, smilkov2017smoothgrad, sundararajan2017axiomatic, selvaraju2017grad, montavon2019layer}. 
It has also been applied in the context of time series~\citep{ismail2020benchmarking, fawaz2019deep}

\noindent \textbf{Application to 
time-series gravitational wave data:} 

Recall the architecture of the ML model: In the case of 
the \texttt{WaveNet} models we use for 
this study~\citep{huerta_nature_ast}, both the input and output 
layers are vectors in 4096 dimensions. The neurons 
in the output layer report the probability of the presence of 
a gravitational wave prior to that neuron, and zero 
otherwise. For example, if the input data contains 
a gravitational wave signal that describes a binary black 
hole whose merger occurs at timestep 3000 (in the 
4096-D input), then the output neurons will report 
values close to 1 for all the neurons prior to  
timestep 3000 and zero otherwise. By computing the 
Jacobian between all the neurons in 
the output and input layer, we arrive at 
a \(4096\times4096\) matrix. Following the work cited above, 
we interpret this matrix as a ``map'' of the sensitivity 
of each of the 4,096 separate classifications to each of 
the 4096 coordinates of the input vector.

\noindent \textbf{Results for a signal of advanced LIGO noise:} 

\begin{figure}[!ht]
    \centerline{
       \includegraphics[scale = 0.45]{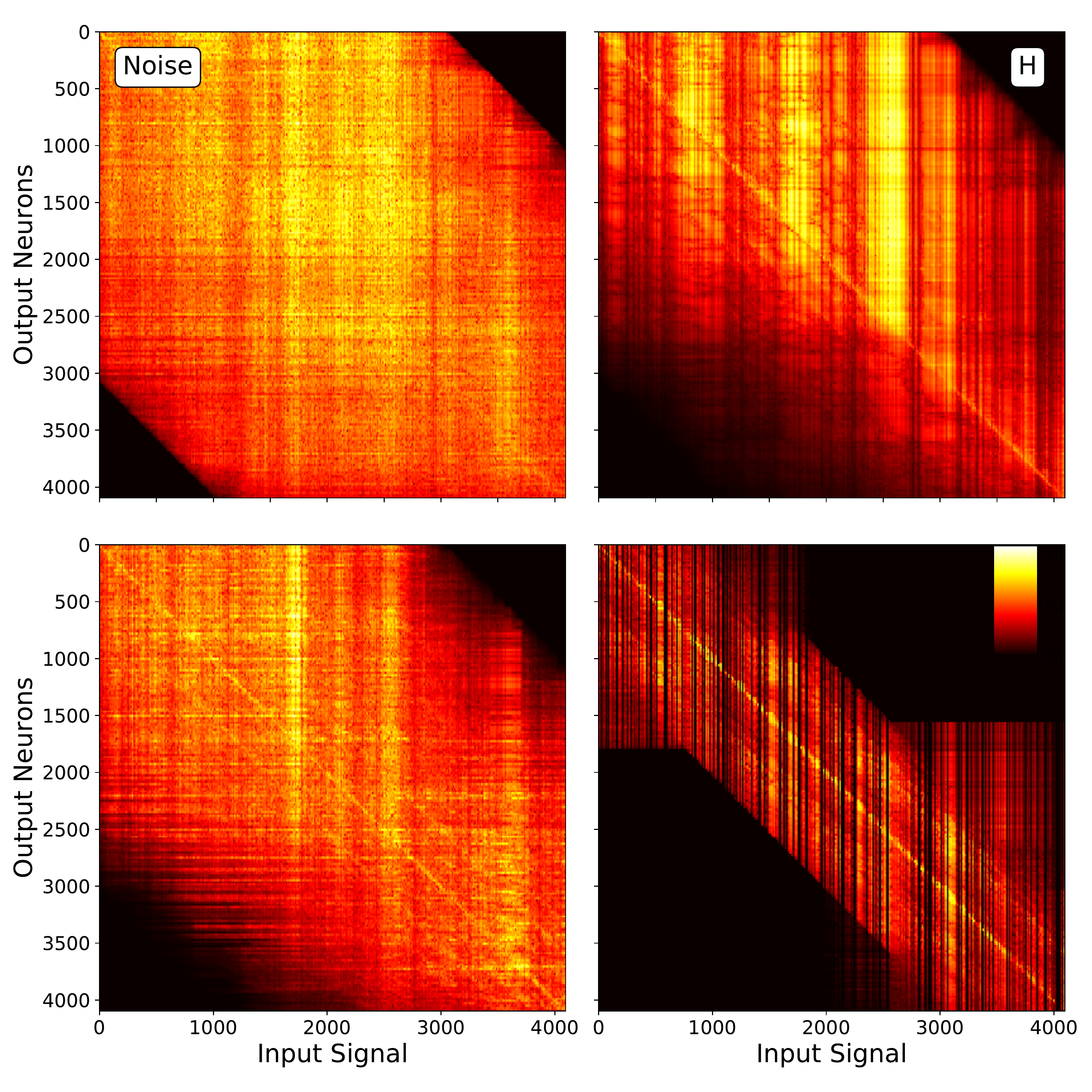}
    } 
\caption{\textbf{Jacobian for advanced LIGO noise inputs of Hanford branch:}
The four panels present the Jacobian, computed using the last 
layer of each of the four models in our AI ensemble, when 
we feed advanced LIGO noise. We identify two recurrent features in these panels: (i) a diagonal 
line, which indicates that the neurons in the final layer are 
always sensitive to the corresponding input neuron in the 
absence of a real signal; (ii) the dark corner regions in the panels are indicative of the restricted field of view of the neurons in the last layer with respect to the input signal. Neurons in the deeper layers have wider field of view.}
\label{fig:jac_noise_H}
\end{figure}

\begin{figure}[!ht]
    \centerline{
       \includegraphics[scale = 0.45]{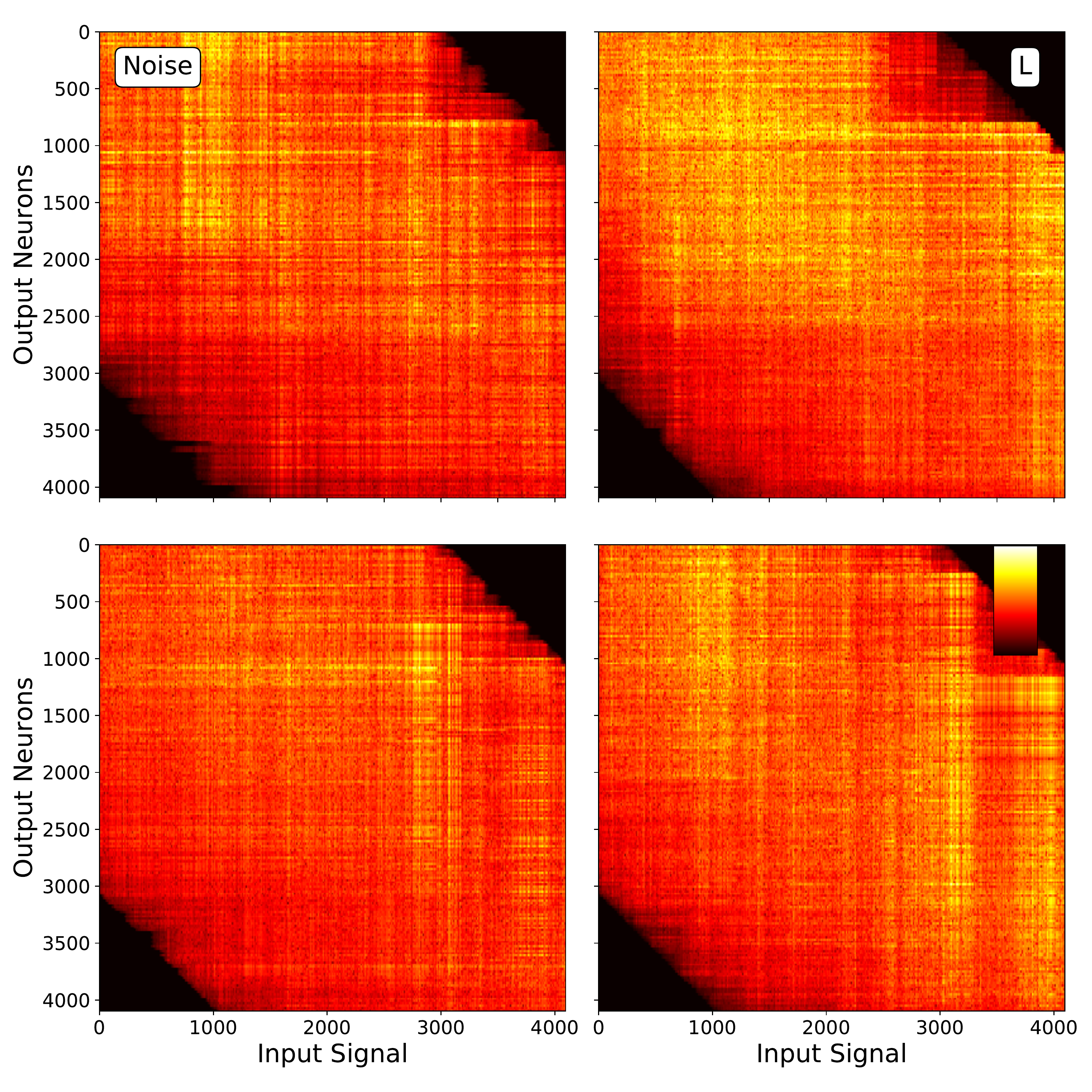}
    } 
\caption{\textbf{Jacobian for advanced LIGO noise inputs:} 
Same as Figure~\ref{fig:jac_noise_H}, but for the Livingston 
branch of each of the four models.}
\label{fig:jac_noise_L}
\end{figure}

As a baseline, we first created a sensitivity for a signal 
containing no merger event, whitened 
advanced LIGO noise inputs. Figures~\ref{fig:jac_noise_H} 
and~\ref{fig:jac_noise_L}
shows four panels that correspond to the Jacobian of each 
of our four ML models for the Hanford and Livingston branches, 
respectively. The $x$-axis in each panel shows 
the input waveform (in this case input noise), while 
the $y$-axis shows the output neuron. 

We notice a clear cut difference between the H and L branches, 
namely, there are large swaths of dormant neurons in the H branch 
for several ML models, while the L branch in our models has
solid activity throughout. This, again, is consistent with 
previous findings in that the L branch is searching for 
features in a global manner, while the H branch has a more narrow 
scope of feature identification.

This sensitivity map offers one more useful observation. Note the prominent dark corners at lower left and upper of the images. These correspond to zero sensitivity. Indeed, a simple calculation from the architecture of the network shows that it is impossible for a final-layer neuron at one end of the signal to be affected by input data at the other end. In other words, the field of view of neurons does not always cover the entire signal.

\begin{figure}[!ht]
    \centerline{
    \includegraphics[scale = 0.45]{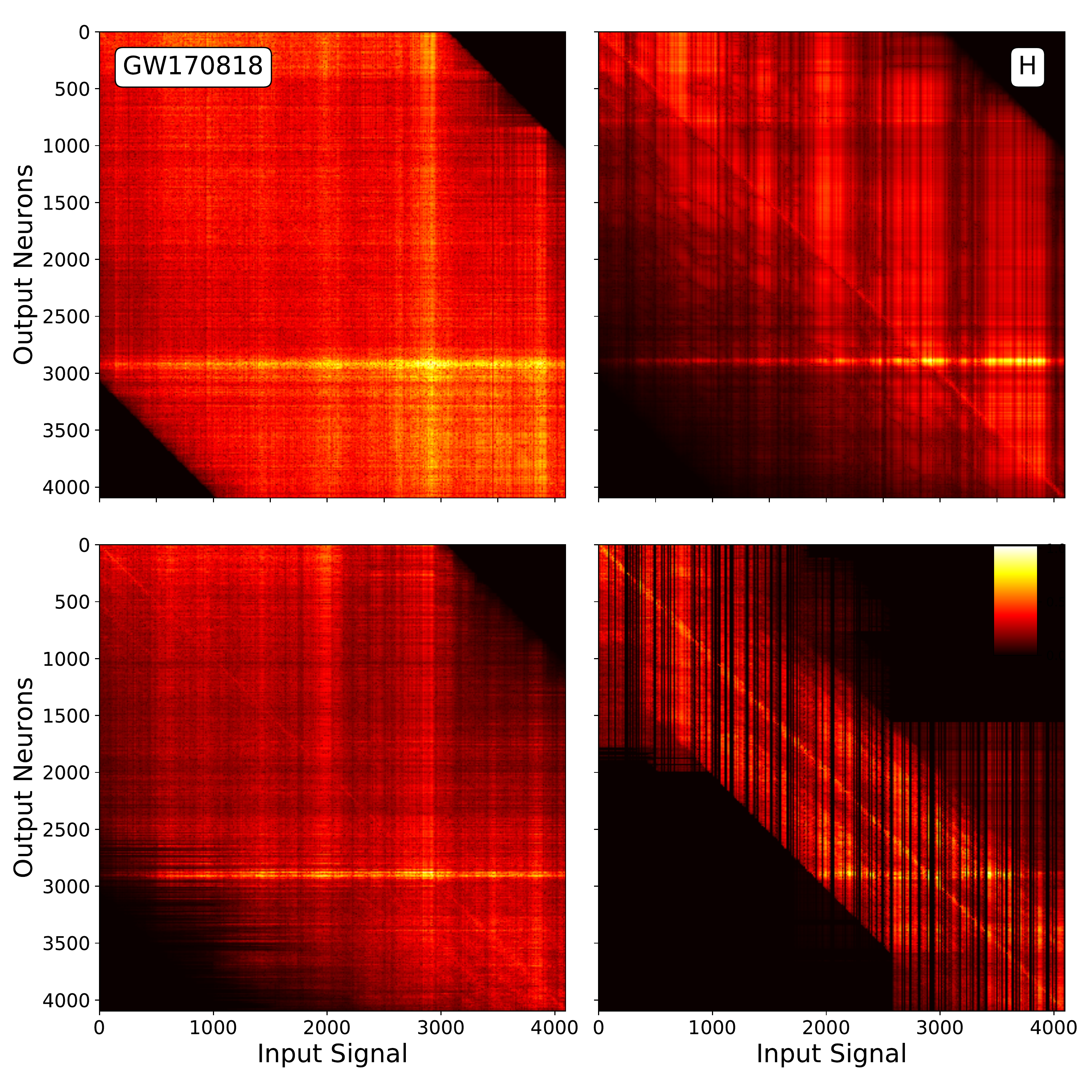}
    }
\caption{\textbf{Jacobian for GW170818:} 
Same as Figure~\ref{fig:jac_noise_H}, but now for the merger event GW170818. These 
results show that when the ML models process advanced LIGO 
data containing real events, output neurons in the last layer 
of these models respond in a distinct manner when they 
identify features that are consistent with the merger 
of binary black holes. These panels contain that information in the sharp vertical and horizontal lines at output neuron 
and timestep 3,000, which coincide with the merger of 
GW170818.}
\label{fig:jac_GW170818_H}
\end{figure}

\begin{figure}[!ht]
    \centerline{
    \includegraphics[scale = 0.45]{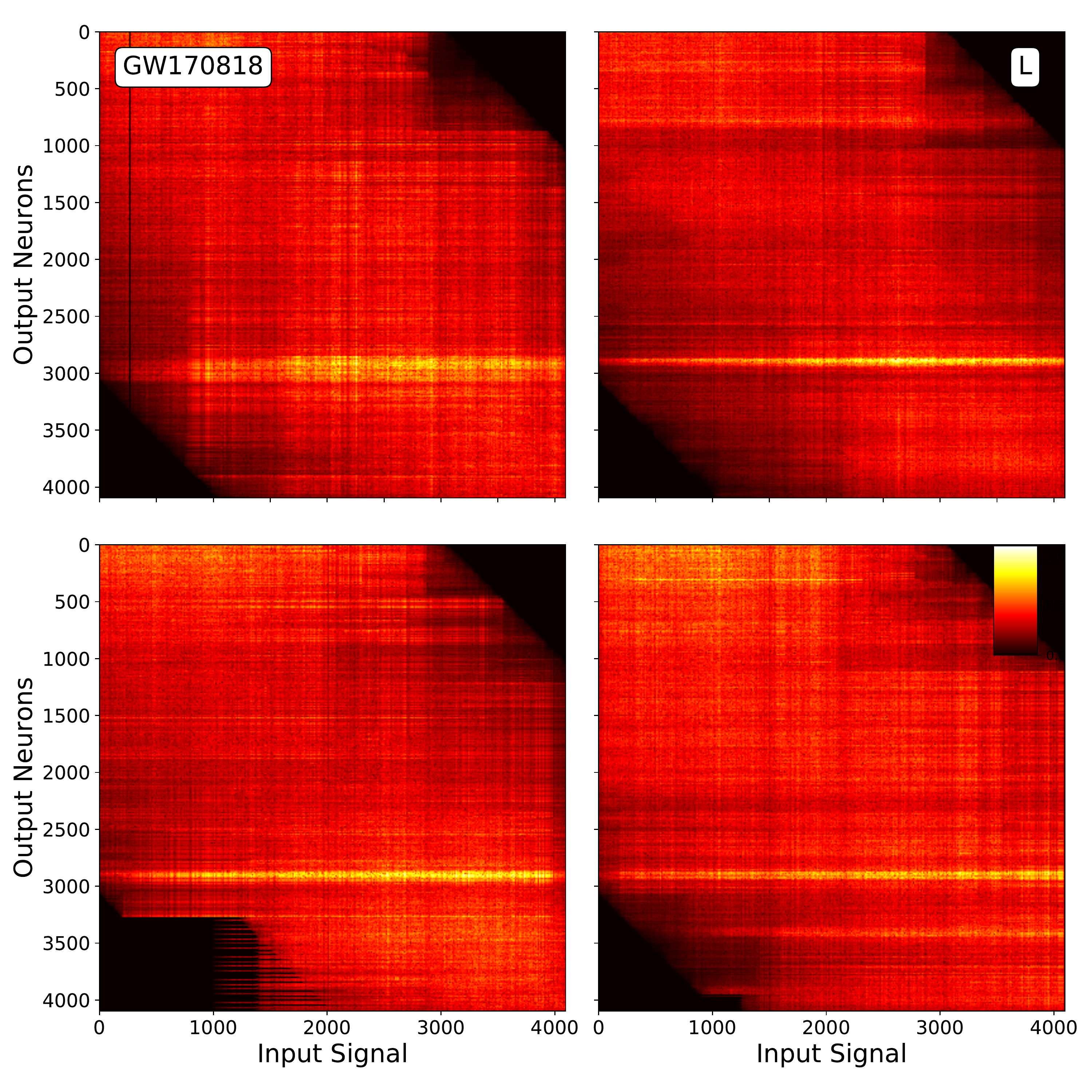}
    }
\caption{\textbf{Jacobian for GW170818:} 
Same as Figure~\ref{fig:jac_GW170818_H}, but for the Livingston branch of each of the four models.}
\label{fig:jac_GW170818_L}
\end{figure}

\noindent \textbf{Results for GW170818:} 

Having analyzed sensitivity for a signal without a merger, we turn to the case of a signal with a verified event: a
binary black hole merger known as GW170818. This event was chosen since it was particularly strong, and thus we may observe a 
sharp response in our ML models.

Comparisons between 
Figures~\ref{fig:jac_noise_H} and \ref{fig:jac_GW170818_H} 
show that our ML models indeed have a visibly different response when 
the input data contains a real signal. 
First, the diagonal line we observe 
in all panels in Figure~\ref{fig:jac_noise_H} is not 
as prominent when the input data contains real signals.

Second, we observe two new lines or bands, one vertical at 
timestep 3,000, and one horizontal at location 3,000. 
The horizontal line indicates that output at the 
location of the merger is sensitive to the entire input 
signal. The vertical line about the location 3,000 suggests 
that all the neurons in the last layer are sensitive 
to information residing at timestep 3,000 in the input 
signal, meaning all the neurons in the last layer are 
sensitive to the merger event.
We observe similar patterns when we make 
pair-wise comparisons of Figures\ref{fig:jac_noise_L} and \ref{fig:jac_GW170818_L}, except we 
do not see a clear vertical line in Figure~\ref{fig:jac_GW170818_L}. 
On the other hand, what remains the same is that 
the Hanford and Livingston branches respond differently to 
real gravitational wave signals. The Livingston branch seems to search for 
global features in the input data, while the Hanford branch appears to focus on 
localized features of the merger event.

To rule out the possibility that the particular timestep 
we chose was somehow special, we conducted additional 
experiments in which we manually shifted the location of 
the merger event, using a different start time to clip 
the advanced LIGO segment 
containing this signal and then feeding it into our models. 
We observed that the time shift we used to clip this input 
data segment coincided with the actual location 
of these horizontal and vertical lines. We also used several 
synthetic signals that we injected in advanced LIGO noise 
and repeated these experiments several times. In all these 
experiments we observed a strikingly similar 
feature response.

\section{Discussion}
\label{sec:end}

The results of the three experiments suggest several hypotheses about how the GW ensemble functions. We describe how individual units in the network appear to respond in an intuitive way to aspects of the incoming signal. At the architectural level, the two-branched form of the network seems to lead to an unexpected separation of concerns: one branch seems to focus on more local features, another on denoising and global features.

\subsection{Signals and individual neural responses}

At a basic level, all three visualization techniques reveal that these neural networks may be less mysterious than they appear. All of the activation layers of the network 
react differently when a signal is present versus 
pure noise. When there is no signal in the input data, 
the neurons in the last layer are mostly sensitive 
to the corresponding neuron in the input layer. In the presence of a signal in the data, the 
neuron in the last layer around the merger event 
becomes sensitive to the entire input time-series data. 

Furthermore, the type of signal matters. The network reacts differently to 
stellar mass vs heavier binary black hole mergers. 
We consistently noticed that more neurons are 
actively involved in detection of lighter systems, 
as opposed to their heavier counterparts. Perhaps if a signal is harder to detect, larger parts of the neural 
net will be involved in the classification task. Finally, each of the four models react differently to the input data, which indicates each iteration of training of the model has led to slight differences in how a net has learned to tell apart signals from noise anomalies.

Activation maximization experiments suggest that channels and individual neurons are sensitive to ``template'' signals that resemble part or whole waves, with a distinct visual similarity to training data examples. Moreover, early layers 
are activated by the presence of a simple spike, 
while deeper layers are activated by more 
complicated signal. Visualizing activations and sensitivity maps directly shows that the network seems to find the moment of a black hole merger especially salient. All three experiments indicate a progression from very local to more global feature extraction. These results seem intuitive, and match similar results from computer vision networks.

\subsection{Branch specialization}

Beyond these basic observations, however, our visualizations point to an unexpected finding, painting a scenario in which 
the two branches of our neural network 
have specialized for somewhat different tasks. The Livingston branch seems to focus on global features of waveform 
signals. Synthetic signals designed to optimize the Livingston response generally were much noisier than equivalent signals for the Hanford branch, and seemed to span a wider time scale. We hypothesize the Livingston branch may have a special role in identifying 
global, longer time-scale features of waveform signals. 
Meanwhile, the Hanford branch appeared to focus on shorter-time events that were smoother. Working in tandem, these branches 
learn to identify key features that define the physics 
of binary black hole mergers and which are not 
present in other types of noise anomalies. 

This difference between branches raises a number of questions. Branch specialization has been observed to occur spontaneously\citep{voss2021branch} simply due to architectural separation within a neural network. In the case of the GW ensemble, however, there is the additional factor that the two branches are given signals that, although similar, have slightly different noise characteristics. Understanding the relation between the differences in training data and the resulting specialization seems like an important area for future research. Furthermore, understanding the emergent differences in processing may help in evaluating future versions of the network, or even choosing which trained networks to include in an ensemble.

\section{Conclusion}

\noindent We draw three general conclusions from our interpretability experiments. First, there is value to this type of translational research: although many of the techniques we have applied were first developed in other domains, they appear to shed light in our context as well. Moreover, all three techniques paint a consistent picture of a network whose branches are performing different types of computations, and which responds to increasingly global features as layer depth increases.

The second conclusion is that there is significant room for new research based on these initial findings. One natural direction is looking for ways to modify the basic network architecture. For example, in the current models the Jacobian investigation brings up the fact that some units in the final layer are not sensitive to the entire field; however, it would be straightforward to change the architecture to avoid this. At a deeper level, the activation visualizations indicate that in some cases a large portion of the network seems to respond similarly to signals with and without an event; a close examination of this phenomenon might suggest ways to compress or prune the network. 

Finally, our visualizations raise additional avenues for interpretation. In particular, we have hypothesized that the specialization of the two branches relates at least partly to the difference in baseline noise levels in the two detectors. Understanding how differences in data distribution might affect spontaneous branch specializing is an appealing question. From a practical standpoint, it would be useful to understand whether this type of specialization contributes positively to performance or has a neutral or negative effect.


\section{Acknowledgements}
\label{ack}
\noindent We are thankful to David Bau for careful reading of our paper before submission. A.K. and E.A.H. gratefully acknowledge National 
Science Foundation (NSF) awards OAC-1931561 and 
OAC-1934757. E.A.H. gratefully acknowledges the Innovative and 
Novel Computational Impact on Theory and Experiment 
project `Multi-Messenger Astrophysics 
at Extreme Scale in Summit'. This material is 
based upon work supported by Laboratory Directed 
Research and Development (LDRD) funding from Argonne National 
Laboratory, provided by the Director, Office of Science, of the 
U.S. Department of Energy under Contract No. DE-AC02-06CH11357. 
This research used resources of the Argonne 
Leadership Computing Facility, which is a DOE Office of 
Science User Facility supported under Contract 
DE-AC02-06CH11357. This research used resources of the 
Oak Ridge Leadership Computing Facility, 
which is a DOE Office of Science User Facility 
supported under contract no. DE-AC05-00OR22725. 
This work utilized resources supported by the 
NSF's Major Research Instrumentation program, 
the HAL cluster (grant no. OAC-1725729), 
as well as the University of Illinois at 
Urbana-Champaign. We thank \texttt{NVIDIA} for 
their continued support. 

\bibliographystyle{plainnat}
\bibliography{ML_GW}

\end{document}